# MONOPOLE ANNIHILATION AND HIGHEST ENERGY COSMIC RAYS

P. Bhattacharjee[a,b] and G. Sigl[c,d]

[a] *Isaac Newton Institute, University of Cambridge, 20 Clarkson Road, Cambridge CB3 0EH, U.K.*

[b] *Indian Institute of Astrophysics Sarjapur Road, Koramangala, Bangalore 560 034, INDIA*[1]

[c] *Department of Astronomy & Astrophysics Enrico Fermi Institute, The University of Chicago, Chicago, IL 60637-1433*

[d] *NASA/Fermilab Astrophysics Center Fermi National Accelerator Laboratory, Batavia, IL 60510-0500*

## ABSTRACT

Cosmic rays with energies exceeding $10^{20}$ eV have been detected. The origin of these highest energy cosmic rays remains unknown. Established astrophysical acceleration mechanisms encounter severe difficulties in accelerating particles to these energies. Alternative scenarios where these particles are created by the decay of cosmic topological defects have been suggested in literature. In this paper we study the possibility of producing the highest energy cosmic rays through a process that involves formation of metastable magnetic monopole-antimonopole bound states and their subsequent collapse. The annihilation of the heavy monopole-antimonopole pairs constituting the monopolonia can produce energetic nucleons, gamma rays and neutrinos whose expected flux we estimate and discuss in relation to experimental data so far available. The monopoles we consider are the ones that could be produced in the early universe during a phase transition at the grand unification energy scale. We find that observable cosmic ray fluxes can be produced with monopole abundances compatible with present bounds.

[1] Permanent Address

# 1 Introduction

The physics and astrophysics of cosmic rays (CR) of ultrahigh energy (UHE) (i.e. with energy above about $10^{18}$ eV) constitute a subject of much intense research [1, 2, 3] in recent times both in terms of new experiments as well as new theories. UHE CR with energies exceeding $10^{20}$ eV have been detected [4, 5, 6, 7, 8, 9, 10]. The Haverah Park experiment [4] reported several events with energies near and slightly above $10^{20}$ eV. The world's highest energy CR event detected recently by the Fly's Eye experiment [7, 8] has an energy $\sim 3\times10^{20}$ eV. The event of energy $\sim 1.1\times10^{20}$ eV recorded by the Yakutsk experiment [5, 6] has almost the same arrival direction as that of the Fly's Eye event. More recently, the AGASA experiment [9] has also reported an event with energy (1.7–2.6)$\times10^{20}$ eV [10].

The existence of UHE CR, especially, the highest energy cosmic rays (HECR) (i.e. with energy above $10^{20}$ eV) poses serious challenge [11, 12, 13, 14, 15, 16] for conventional astrophysical acceleration mechanisms [17] that attempt to explain the origin of these particles in terms of acceleration in special astrophysical sites like supernova shocks, pulsar magnetospheres, galactic wind termination shocks or relativistic shocks associated with active galactic nuclei and radio galaxies. In this last case acceleration up to around $10^{21}$ eV seems to be possible by stretching the "reasonable" values for the shock size and the magnetic field strength at the shock somewhat [18]. However, at least for the highest energy Fly's Eye and Yakutsk events mentioned above, as also for the more recently detected AGASA event [10], there seem to be no suitable extragalactic objects such as AGNs or rich galaxy clusters near the observed arrival directions and within a maximum distance of about 50 Mpc, this upper limit on the possible source distance being set by considerations of energy loss during propagation [13, 14, 15, 16]. Thus it is difficult to associate these highest energy events with any known astrophysical sources.

The difficulties encountered by conventional acceleration mechanisms in accelerating particles to the highest observed energies have motivated recent suggestions [19, 20, 21, 15] that the underlying production mechanism of HECR could instead be of a non-acceleration nature, namely the decay of supermassive elementary "X" particles related to Grand Unified Theories (GUTs). Sources of such particles today could be topological defects (TDs) [22] formed in the early universe during phase transitions associated with spontaneous breaking of symmetries implemented in these GUTs. This is because TDs like cosmic strings, domain walls, superconducting cosmic strings and magnetic monopoles are topologically stable but nevertheless can release part of their energy in the form of these X particles due to physical processes like collapse or annihilation. The X particles, with masses of the typical GUT scale which is generally much higher than $10^{21}$ eV, decay into leptons and quarks, the latter ones finally hadronizing into jets of hadrons and giving rise to HECRs. In this scenario, the observed HECR are due to collapse or annihilation of TDs at relatively "close" distances ($\lesssim 50$ Mpc) from earth.

The predicted spectra of UHE particles in the TD scenario are determined essentially by the physics of hadronization of quarks, i.e., by Quantum Chromodynamics (QCD). In this sense the shapes of the spectra are universal (i.e., independent of the specific process involving any specific kind of TDs), especially at the highest energies where cosmological evolutionary effects are negligible (except for neutrinos). The overall contribution of different TD processes to the UHE cosmic ray flux would, however, be different.

Among the various kinds of possible TDs, the case of cosmic strings is perhaps the one that has been studied most extensively in terms of their formation and subsequent evolution, principally because they provide an attractive theory of formation of galaxies and large-scale structure in the universe. It was, therefore, natural to investigate first the various possible UHE CR producing processes involving cosmic strings. Almost all the mechanisms studied so far in this connection [20, 23, 24] involve closed loops of cosmic strings. It turns out that cosmic strings can give measurable contribution to UHE CR flux only if there is a mechanism by which a fraction $\sim 10^{-5}$ of the energy of all closed loops of cosmic strings chopped off the long (i.e., larger than horizon size) segments of strings at any time is dissipated in the form of X particles on a time scale much smaller than the time scale of energy loss of these loops through gravitational radiation. One process which would satisfy this criterion involves the "completely" collapsing class of loops [20] or loops which undergo multiple self-intersections and break up into a large number of small loops (rather than a small number of large loops) within one period of oscillation of these loops. These loops are, however, rather special in the sense that they have to be fine-tuned to one of the collapsing configurations to within a length scale of the order of the width of the string, which is a microscopic scale of order $\sim 10^{-29}$ cm for GUT scale cosmic strings. Gill and Kibble [25] have recently argued that these "smooth" loops are unlikely to be formed in any significant number except possibly at very early times. To explain HECR, however, the collapse of the loops must occur in relatively recent epochs. Gill and Kibble have, therefore, argued that processes involving *ordinary* cosmic strings are unlikely to yield a measurable flux of UHE CR. It is, however, conceivable [26] that hybrid systems of TDs such as light domain walls bounded by GUT scale cosmic string loops could form in the early universe; the domain walls in these systems could aid the complete collapse of the string loops that form the boundaries of these walls. It is also conceivable [27] that gravitational radiation back reaction effects may smooth out higher frequency wiggles leaving only the lowest-frequency mode waves on the string loops making them collapse completely [28] such that a significant amount of UHE CR could be produced. These possibilities, however, remain to be studied in detail.

Another potential TD-source of HECR are the saturated superconducting cosmic string (SCS) loops [29, 21]. If and when an SCS loop achieves a certain saturation current [29] the massive charge carriers that carry the electric current on the string are ejected from the string. The subsequent decay of these massive charge carriers can produce UHE particles [21]. There are, however, several issues in this context that remain to be settled. The cosmological evolution of current-carrying SCS is much more complicated than evolution of "ordinary" cosmic strings. In particular, it is possible [30] that SCS loops may never achieve the saturation current at all. Another issue of debate [31, 21] is whether or not the charge carriers, assuming they are ejected from the string, can get out of the immediate vicinity of the string before decaying. The decay of the massive charge carriers must occur outside the region of strong magnetic fields surrounding the strings; otherwise, the energetic decay products would rapidly lose energy by interacting with the strong magnetic field and so would not survive as UHE particles.

These issues certainly remain to be studied in detail. At the same time, it is worthwhile, in our opinion, to study the possibility of production of measurable fluxes of UHE cosmic rays by other possible processes involving other kinds of TDs. With this motivation, we investigate in this paper another possible TD-process of generating UHE cosmic rays, namely the collapse of relic magnetic monopole-antimonopole bound states (monopolonium), first suggested by Hill [32].

Magnetic monopoles (simply "monopoles" hereafter) are one kind of TD solutions in spontaneously broken non-abelian gauge theories that are allowed in essentially all GUT models. If $m_X$ denotes the mass of a typical gauge boson in a GUT, then the monopole mass $m_M$ is given by $m_M \sim \alpha_X^{-1} m_X$, where $\alpha_X$ is the dimensionless "unified" gauge coupling strength at the energy scale $m_X$ in the GUT model under consideration. The core of a "GUT monopole" has a radius $\sim m_X^{-1}$. Formation of massive monopoles in the early universe, through Kibble mechanism [22], is inevitable in most GUT models, and leads to the well-known "monopole problem" (see, for a review, Ref. [33]). The "problem", to remind the reader, refers to the "standard model" prediction of relic abundance of monopoles which, on various empirical grounds, is unacceptably large. (Here, by "standard model" prediction we mean prediction made within the context of the simplest GUT models and the standard Big-Bang cosmology). Several mechanisms including cosmic inflation [34] have been proposed for reducing the relic monopole abundance to acceptable levels (see, for a review, [33]). For instance, interesting relic abundances could have been produced thermally during reheating after the universe has gone through an inflationary phase. However, we will not discuss those mechanisms in this paper. Instead, we will simply assume that monopoles exist in the universe at a level of abundance compatible with known experimental [35] and phenomenological [36, 37] upper bounds.

Since monopoles are topologically stable, the only way of getting rid of monopoles ($M$s) is to make them annihilate with antimonopoles ($\bar{M}$s). The "standard" way (see, e.g., Refs. [38, 39]) of achieving this relies upon mechanisms to capture $M - \bar{M}$ pairs in metastable bound states which spiral in and finally collapse resulting in annihilation of the $M$s and the $\bar{M}$s that have been captured in bound states. However, as is well known, this typically is a slow process and fails in "solving" the monopole problem, i.e., the $M$s and $\bar{M}$s do not annihilate fast and early enough for the Universe to avoid being monopole dominated. *If*, however, the monopole problem is "solved" to start with (by some mechanism which we do not concern ourselves with in this paper), i.e., if the universe is never monopole dominated to start with, which we shall assume to be the case in this paper, then the late annihilation of the monopoles is, in fact, precisely the mechanism that we need from the point of view of generating UHE cosmic rays which, as we know, must be produced only in the contemporary cosmic epoch (i.e., at low redshifts); the potential UHE cosmic ray particles resulting from $M - \bar{M}$ annihilation occuring in the earlier epochs would thermalize by interacting with the dense background medium and hence would not survive as UHE particles. So monopoles formed at a GUT-scale phase transition in the early universe and annihilating in the recent epochs provide us with an attractive scenario of production of UHE cosmic rays provided they exist in the Universe in sufficient numbers. The main aim of the present paper is, in fact, to try to estimate the monopole abundance required in order to generate enough UHE CR flux as observed. Furthermore, we discuss the predicted particle spectrum in relation to experimental data so far available.

We organize the rest of the paper as follows: In Sec. 2 we describe the relevant framework of monopolonium physics. In Sec. 3 we calculate the contribution of monopolonium collapse to the UHE CR flux in terms of the model parameters and estimate the abundance of monopolonia (relative to that of monopoles at the relevant time of formation of the monopolonia) required to produce enough flux of the highest energy cosmic rays as observed. In section 4, we then study, following the original analysis of Hill [32], whether the required monopolonia abundance established in section 3 are realizable. Finally, in Sec. 5 we summarize our findings.

Wherever appropriate we shall use natural units with $\hbar = c = k_B = 1$, $k_B$ being the Boltzmann constant. Quite often, however, we keep one or more of these quantities explicitly in some formulae for convenience, which should be clear from the context. Also, we assume a "flat" ($\Omega_0 = 1$) universe with a Hubble constant $H_0 = 100\, h\, \mathrm{km\,sec^{-1}\,Mpc^{-1}}$ with h=0.75 taken in the actual numerical calculations.

# 2 Monopolonium Physics: A Brief Introduction

Monopolonium [32] is a possible bound state of a magnetic monopole and an antimonopole. Let us first consider, classically, a monopole ($M$) and an antimonopole ($\bar{M}$) separated by a distance $r$ and bound in a circular orbit around each other. The non-relativistic energy of the system can be written as

$$E = \frac{1}{2}\tilde{m}_M \omega^2 r^2 - g_m^2/r\,, \tag{1}$$

and since

$$\frac{g_m^2}{r} = \tilde{m}_M \omega^2 r^2\,, \tag{2}$$

we have

$$E = -\frac{1}{2} g_m^2/r\,, \tag{3}$$

where $\tilde{m}_M = m_M/2$ is the reduced mass, $\omega$ is the orbital angular frequency, and $g_m$ is the magnetic charge of the monopole. The Dirac quantization condition relating the magnetic charge $g_m$ and electric charge $e$ is (keeping $\hbar$ and $c$) [40, 41]

$$\frac{e g_m}{\hbar c} = \frac{N}{2}\,, \tag{4}$$

where the integer $N$ is the "monopole number". For $N = 1$ monopoles, the "magnetic" fine-structure constant $\alpha_m$ is given by

$$\alpha_m \equiv \frac{g_m^2}{\hbar c} = \frac{1}{4}\frac{1}{\alpha_e} \simeq \frac{137}{4} = 34.25\,, \tag{5}$$

where $\alpha_e \equiv e^2/(\hbar c) \simeq 1/137$ is the "electric" fine-structure constant. We will ignore the "running" of the coupling strengths $\alpha_e$ and $\alpha_m$.

"Classical" monopolonium is, of course, unstable (like the "classical" hydrogen atom). A Bohr model of monopolonium can, however, be constructed. In the Bohr model, not all values of $r$ are allowed. Monopolonium can exist only in certain discrete states characterized by the principal quantum number $n$ given by

$$r = n^2 a_m^B \,, \tag{6}$$

where $n = 1, 2, 3, \ldots$, and

$$a_m^B \equiv \frac{\hbar^2}{\tilde{m}_M g_m^2} = 8\alpha_e \left(\frac{\hbar}{m_M c}\right) = 8\alpha_e \alpha_X \left(\frac{\hbar}{m_X c}\right) \tag{7}$$

is the Bohr radius of the monopolonium state. In writing Eq. (7) we have used Eq. (5). Now Eq. (3) can be written as

$$E = -\frac{R_m}{n^2} \,, \tag{8}$$

where $R_m$ is the effective "Rydberg constant" for monopolonium, and is given by

$$R_m = \frac{1}{2} \tilde{m}_M \frac{g_m^4}{\hbar^2} \simeq 2.5 \times 10^{16} \, \alpha_m^2 \left(\frac{m_M}{10^{17}\,\mathrm{GeV}}\right) \mathrm{GeV} \,. \tag{9}$$

(Compare with this the Rydberg constant for the hydrogen atom, 13.6 eV).

Note from Eq. (7) that $a_m^B \ll \hbar/(m_M c) < \hbar/(m_X c)$. In other words, the monopolonium ground state ($n = 1$) is one in which the cores of the monopole and the antimonopole overlap strongly. Clearly, then, monopolonium does not really exist in the $n = 1$ state because the $M$ and the $\bar{M}$ would annihilate each other. Presumably, when monopolonium is formed in the early universe, the $M$ and the $\bar{M}$ capture each other in a state with $n \gg 1$. This state then undergoes a series of transitions to tighter and tighter bound states by emitting initially photons and subsequently gluons, Z bosons, and finally the GUT X bosons. At some stage during the collapsing process, the cores of the $M$ and the $\bar{M}$ begin to overlap (when $r \sim 2/m_X$, i.e., when $n$ reduces to $\sim [m_M/(4\alpha_e m_X)]^{1/2}$) and subsequently annihilate each other. When this happens, the bound state is destroyed and the energy contained in the system is released in the form of various particles. We are interested in estimating the flux of UHE CR produced by these final $M - \bar{M}$ annihilation events associated with collapsing monopolonia in the universe. Actually, the gluons emitted by Larmor radiation during the collapsing process as well as the quarks from the decay of Z-bosons radiated by monopolonia will also hadronize and produce energetic hadrons which will contribute to the total CR flux. We will, however, restrict ourselves here to estimating the CR flux due to the final $M - \bar{M}$ annihilation events discussed above. Therefore, what we will estimate here will be a lower limit to the total possible contribution to the UHE CR flux from monopolonia in the universe.

The monopolonium lifetime $\tau$ can be calculated [32] by using the dipole radiation formula

$$\frac{dE}{dt} = -\frac{64}{3}\frac{E^4}{g_m^2 m_M^2 c^3}\,. \tag{10}$$

A monopolonium state formed at a time $t_f$ collapses (i.e., $r$ becomes $\sim m_X^{-1}$) at the time $t_c$ which can be calculated by integrating Eq. (10), which gives,

$$E^{-3}(t_c) - E^{-3}(t_f) = \frac{64}{g_m^2 m_M^2 c^3}\cdot(t_c - t_f)\,, \tag{11}$$

where $E(t_f) = -g_m^2/(2r_f)$, $r_f$ being the radius of the monopolonium at the time of its formation, and $E(t_c) = -g_m^2/(2r_X)$ with $r_X \sim m_X^{-1}$.

In the situation relevant for our case, $r_f \gg r_X \sim 10^{-29}$cm. The lifetime $\tau \equiv t_c - t_f$ is, therefore, given by

$$\tau = \frac{m_M^2 c r_f^3}{8\alpha_m^2 \hbar^2}\,. \tag{12}$$

Note that $\tau \propto r_f^3$. Thus, for example, $\tau \sim 40\,$days if $r_f \sim 1\,$fm, whereas $\tau \sim 10^{11}\,$yr if $r_f \sim 1$ nm. In other words, depending on the initial radius of the bound state at its formation, some of the monopolonia formed in the early universe could be surviving in the universe today and some would have collapsed in the recent epochs.

An operational definition of "formation" of monopolonium states in the universe can be taken as follows. At any time $t$ when the temperature of the universe is $T$, $M-\bar{M}$ bound states with binding energy $E_b = \eta T$ (where $\eta \gtrsim O(1)$ is an unknown parameter at this stage) "freeze out" or "form" and start to collapse. The parameter $\eta$ incorporates the requirement that the formed monopolonia not be thermally dissociated. (We are, of course, implicitly assuming here that monopoles interact efficiently enough with the background plasma of thermal electrons and photons at the relevant times of formation of the bound states; see below). The abundance of monopolonium formed can in principle be estimated if a specific mechanism of formation is given. We will not go into the discussion of any specific monopolonium formation mechanism here. Instead, we will first try to estimate, on a phenomenological basis, the monopolonium abundance (relative to a given abundance of monopoles) required for the scenario to yield measurable flux of UHE cosmic rays. In section 4, we will estimate the relative monopolonium abundance within the framework of a Saha formalism applied to a system of $M$s, $\bar{M}$s and $M-\bar{M}$s in thermal equilibrium, and compare the abundance so obtained with the phenomenologically required estimate. We shall see that within the framework of the thermal equilibrium analysis the relative monopolonium abundance is essentially parametrized by the parameter $\eta$ mentioned above.

We are interested in monopolonium states that are collapsing in the present epoch, i.e., $t_c \gg t_{eq}$, where $t_{eq}$ is the time of equality of radiation and matter energy densities. On the other hand, with the operational definition of formation of monopolonia mentioned above and the life-time formula, Eq. (12), we have, $t_f \ll t_{eq}$, and, therefore,

$$E_{bf} \equiv \eta T_f \approx \left(\frac{64 t_c}{g_m^2 m_M^2 c^3}\right)^{-1/3}\,, \tag{13}$$

where

$$T_f = 1.56 \times 10^{-3}\, g_*^{-1/4} \left(\frac{t_f}{\text{sec}}\right)^{-1/2} \text{GeV}\,. \tag{14}$$

In Eq. (14) $g_*$ is the total number of effective relativistic degrees of freedom determining the energy density of radiation in a radiation dominated universe [33]. Eq. (13) reveals that monopolonium states undergoing final collapse today must have been formed with a binding energy $E_{bf} \simeq 0.5\,(m_M/10^{16}\,\text{GeV})^{2/3}$ MeV, corresponding to a temperature of the universe at the time of formation, $T_f$, also of this order, i.e., around the epoch of primordial nucleosynthesis. Comparing the typical time scale for monopole-plasma energy exchange [38], $\tau_s \sim 6.58 \times 10^{-3}\, g_*^{-1/2}\,(m_M/10^{16}\,\text{GeV})\,(\text{MeV}/T)^2$ sec, with the Hubble time (age), $t_u \sim 2.42\, g_*^{-1/2}\,(\text{MeV}/T)^2$ sec, of the radiation (i.e., relativistic plasma) dominated universe, we see that $\tau_s < t_u$ as long as $m_M < 3.7 \times 10^{18}$ GeV. In other words, the monopoles may be assumed to be in thermal equilibrium with the electron-positron-photon plasma at the time of formation of the monopolonium states relevant for our considerations. However, the $e^+e^-$ annihilations at $T \simeq 0.3$ MeV [33] significantly reduces the effectiveness of monopole-plasma scatterings in maintaining thermal equilibrium of the monopoles. Thus, while the relevant bound states may be *assumed* to be formed when the monopoles are still in thermal equilibrium (although only marginally so if we take into account the requirement that $\eta > 1$ so that $T_f < E_{bf}$), their subsequent "spiraling in" and collapse essentially occurs in a situation in which the monopoles are effectively decoupled from the background medium. This justifies, albeit *a posteriori*, our use of the "vacuum" dipole radiation formula, Eq. (10), in calculating the lifetime of the relevant monopolonia at least at the level of approximation adopted in this paper.

We shall use Eqs. (13) and (14) for calculating $t_f$ for a given value of $t_c$ or vice versa.

# 3 Contribution of Monopolonium Collapse to UHE CR

## 3.1 The Rate of X Particle Production

The number of X particles released due to annihilation of $M$ and $\bar{M}$ constituting monopolonium can be simply taken to be $2m_M/m_X$. The number density of X particles released per unit time, $dn_X/dt_i$, due to collapsing monopolonia in the universe at any time $t_i \equiv t_c$ is then given by

$$\frac{dn_X}{dt_i} = \frac{dn^c_{M\bar{M}}}{dt_i}\frac{2m_M}{m_X}\,, \tag{15}$$

where

$$\frac{dn^c_{M\bar{M}}}{dt_i} = \frac{dn^f_{M\bar{M}}}{dE_{bf}}\frac{dE_{bf}}{dt_i}\left(\frac{1+z_i}{1+z_f}\right)^3\,. \tag{16}$$

Here $n^c_{M\bar{M}}$ is the number density of monopolonia collapsing at the time $t_i$ and $n^f_{M\bar{M}}$ is their density at formation. Also $z_i$ and $z_f$ are the redshifts corresponding to the times $t_i$ and $t_f$. The redshift factor in Eq. (16) takes care of the dilution of the monopolonium number density due to expansion of the universe.

### 3.2 Injection Spectra of Nucleons, Gamma Rays and Neutrinos

We now assume that each X-particle decays into a lepton and a quark each of approximate energy $m_X/2$. The quark hadronizes by jet fragmentation and produces nucleons, gamma rays and neutrinos, the latter two from the decay of neutral and charged pions in the hadronic jets. The lepton can also generate further particles by interacting with the background medium. But we expect the hadronic route will generate by far the largest number of particles, and we will concentrate on these.

The spectra of the hadrons in a jet produced by the quark are, in principle, given by QCD. Suitably parametrized QCD motivated hadronic spectra that fit well the data in collider experiments in the GeV–TeV energies have been suggested in the literature [32]. Below, we shall illustrate our results by using the QCD motivated spectra suggested by Hill [32]. It is to be kept in mind, however, that there is a great deal of uncertainty involved in extrapolating the formula that describe the "low" energy data to the extremely high energies as in the present situation. To study the sensitivity of our results to the assumed hadronization spectrum, we will also present the results for injection spectra suggested in Ref. [42] on certain phenomenological grounds.

The injection spectrum, i.e., the number density of particles produced per unit time per unit energy interval, for the species $a=$ nucleons (N), gamma rays ($\gamma$) and neutrinos ($\nu$), can be written as

$$\Phi_a(E_i, t_i) = \frac{dn_X(t_i)}{dt_i} \frac{2}{m_X} \frac{dN_a(x)}{dx}\,, \tag{17}$$

where $x \equiv 2E_i/m_X$, $E_i$ being the energy at injection and $dN_a/dx$ is the effective fragmentation function describing the production of the particles of species $a$ from the original quark. We will consider the following two cases for the fragmentation function:

(1) *QCD-motivated injection spectra*

In this case, the *total* hadronic fragmentation spectrum $dN_h/dx$ is taken in the form [32]

$$\frac{dN_h(x)}{dx} = \begin{cases} \frac{15}{16} x^{-1.5}(1-x)^2 & \text{if } x_0 \le x \le 1 \\ 0 & \text{otherwise} \end{cases}\,, \tag{18}$$

where the lower cutoff $x_0$ is typically taken to correspond to a cut-off energy $\sim 1\,\mathrm{GeV}$. Assuming a nucleon content of $\sim 3\%$ and the rest pions, we can write the fragmentation spectra as [19, 43]

$$\begin{aligned} \frac{dN_N(x)}{dx} &= (0.03)\,\frac{dN_h(x)}{dx}\,, \\ \frac{dN_\gamma(x)}{dx} &= \left(\frac{0.97}{3}\right) 2 \int_x^1 \frac{1}{x'} \frac{dN_h(x')}{dx'} dx'\,, \\ \frac{dN_\nu(x)}{dx} &= \left(0.97 \times \frac{2}{3}\right) 2.343 \int_{2.343x}^1 \frac{1}{x'} \frac{dN_h(x')}{dx'} dx'\,. \end{aligned} \tag{19}$$

The neutrino spectrum above includes only the $(\nu_\mu + \bar{\nu_\mu})$'s resulting from the first-stage of the charged pion decay, i.e., from $\pi^\pm \to \mu^\pm + \nu_\mu\,(\bar{\nu}_\mu)$. The further decay of the $\mu^\pm$, i.e., $\mu^\pm \to e^\pm + \nu_e(\bar{\nu}_e) + \bar{\nu}_\mu(\nu_\mu)$, produces additional neutrinos of approximately the same spectra as those of the "first stage" neutrinos. Thus altogether we may expect to have roughly twice as many muon neutrinos as given by the last line of Eq. (19). For a conservative estimate, however, we shall show our results only for the "first stage" neutrinos mentioned above.

(2) *Phenomenological injection spectra*

Recently, injection spectra somewhat different from the QCD motivated injection spectra described above have been suggested by Chi et al [42] on the following phenomenological grounds. UHE gamma rays as well as protons generate lower energy gamma rays by $\gamma - \gamma_b$ and $p - \gamma_b$ collisions with the photons $(\gamma_b)$ of the background radiation fields. The electromagnetic component of the energy lost by the photons and protons in these collisions cascades down to lower energies by electromagnetic cascading in the universal radio background (URB), the cosmic microwave background (CMBR) and in the infrared background (IRB) (in order of decreasing energy of the propagating photon). The measured flux of extragalactic gamma rays in the 100 MeV energy region [44] provides constraints on the form of the nucleon and gamma ray injection spectra at energies above $\sim 5\times10^{19}\,$eV. Based on these considerations, Chi et al [42] have suggested injection spectra which we describe by the following fragmentation functions:

$$\begin{aligned} \frac{dN_N(x)}{dx} &= A_N\, x^{-1.5}\,, \\ \frac{dN_\gamma(x)}{dx} &= A_\gamma\, x^{-2.4}\,, \\ \frac{dN_\nu(x)}{dx} &= A_\nu\, x^{-2.4}\,, \end{aligned} \tag{20}$$

with

$$A_\nu = A_\gamma \approx 0.01\,, \tag{21}$$

and

$$\frac{A_\gamma}{A_N} \simeq 0.028 \left(\frac{10^{15}\,\mathrm{GeV}}{m_X}\right)^{0.9} . \tag{22}$$

The condition (22) comes from the requirement [42] that the ratio of photon-to-nucleon $(\gamma/N)$ at injection at energy $E = 10^{20}\,$eV, i.e., at $x = 2\times10^{20}\,\mathrm{eV}/m_X$ be $\sim 60$. The spectra (20) are assumed to be valid above a suitable lower cutoff $x_0 \lesssim 10^{-4}$.

## 3.3 The Evolved Spectra

The evolution of the spectra is governed by energy-loss and/or absorption of the particles as they propagate through the extragalactic medium. We are interested in calculating the expected diffuse flux assuming that the monopolonia are distributed uniformly. The general expression for the expected diffuse flux today $(t = t_0)$, i.e., the number of particles crossing per unit area per unit time per unit solid angle per unit energy interval at an energy $E_0$, can be written as [20, 23]

$$j(E_0) = \frac{3}{8\pi}\, ct_0 \int_0^{z_{i,max}(E_0)} dz_i (1+z_i)^{-5.5} \frac{dE_i(E_0, z_i)}{dE_0} \Phi(E_i, z_i)\,, \tag{23}$$

where $z_i$ is the injection redshift corresponding to the injection time $t_i$, $E_i(E_0, z_i)$ is the necessary injection energy, and the maximum injection redshift $z_{i,max}(E_0)$ is determined from the condition $E_i(z_{i,max}, E_0) \leq m_X/2$.

We will use the continuous energy-loss approximation [45, 46] for all particles. The general energy-loss equation in terms of redshift $z$ can be written as [20, 23, 45]

$$\frac{1}{E}\frac{dE}{dz} = \frac{1}{1+z} + \frac{(1+z)^{1/2}}{H_0} \beta[(1+z)E]\,. \tag{24}$$

The first term is due to redshift of the (relativistic) particle energy and the second term describes losses due to interactions with the background medium in terms of the inverse energy-loss time scale $\beta(E) = -(dE/dt)/E$. For a particle observed at the earth with an energy $E_0$ the necessary injection energy $E_i(z_i)$ at redshift $z_i$ can be found by integrating Eq. (24) back from $z = 0$ to $z = z_i$ with $E_i(z = 0) = E_0$.

For nucleons we shall use the expression for $\beta(E)$ derived in Ref. [45]. The important process here is the Greisen-Zatsepin-Kuzmin (GZK) [47] effect in which UHE nucleons above a threshold energy ($\sim 6 \times 10^{19}$ eV) lose energy drastically due to photopion production off the CMBR photons. This gives rise to the onset of a sharp fall ("cutoff") of the extragalactic UHE nucleon flux. The interactions of UHE nucleons with the CMBR also cause secondary gamma rays and neutrinos. We shall neglect their contribution to the total gamma ray and neutrino flux calculated below thus getting lower limits for the fluxes of these particles.

The evolution of the UHE gamma ray spectrum is mainly governed by *absorption* of the UHE photons through $e^+e^-$ pair production on the CMBR and on the URB. Actually, under certain circumstances, the propagating photons give rise to electromagnetic cascades [43]. This results in an increase of the effective penetration length of the UHE gamma rays, which, in turn, has the effect of increasing the final gamma ray flux. The cascading effect, however, depends rather strongly on the strength of the intergalactic magnetic field which is rather uncertain. We will ignore here the cascading effect and consider only the absorption of the UHE photons on the CMBR and URB photons. This will give us a conservative estimate of the final gamma ray flux leading to a conservative estimate of the required relative monopolonium abundance, i.e., the actual required monopolonium abundance should be even lower than what we estimate. We will take the absorption lengths for UHE gamma rays as given in Ref. [43]. The relevant absorption lengths being small compared to the Hubble size of the universe, the cosmological redshift term in Eq. (24) is essentially immaterial. Thus for gamma rays we consider only absorption and no energy-loss of the propagating gamma rays.

For UHE neutrinos the dominant process relevant for evolution of the spectrum is the absorption [48, 49] of neutrinos through the process $\nu + \bar{\nu}_b \rightarrow f\bar{f}$, where $f = e, \mu, \tau, u, d, s, c,$ etc., and $\nu_b$ represents the thermal background neutrinos (which have a present temperature $\sim 1.9$ K). It has been pointed out [50] that UHE neutrinos would also generate a neutrino cascade effectively increasing the "neutrino horizon" of the universe, which, in turn, has the effect of increasing the final flux of neutrinos. For example, for an X-particle mass $m_X = 10^{16}$ GeV the flux at $10^{19}$ eV is increased by about a factor 6 [50]. Again, since we are interested in a conservative estimate of the flux, we will ignore this cascading effect here. We will take the absorption redshift for UHE neutrinos as given in Ref. [19].

### 3.4 Required Monopolonium Abundance: An Approximate Analytical Estimate

With the knowledge of energy attenuation lengths and absorption lengths for various particles as described above, we can calculate the expected flux of particles at earth. We will discuss the results in detail in the next subsection. Here we present an approximate analytical estimate of the required monopolonium abundance (relative to a given abundance of monopoles at the relevant time of monopolonium formation) that would yield sufficient flux of UHE cosmic rays as observed. We will do this by estimating the expected UHE gamma ray flux and matching it with the observed flux at a given energy.

We define the relative monopolonium abundance at formation, $\xi_f$, for the monopolonia collapsing in the present epoch, as

$$\xi_f \equiv \frac{1}{n_M^f} \int \frac{dn_{M\bar{M}}^f}{dE_{bf}} dE_{bf} \,, \tag{25}$$

where the integral is understood to be taken over a range of binding energies at formation for which the corresponding monopolonia would be collapsing within, say, one Hubble time period ($\sim t_c$) at the time of collapse $t_c$ which for the CR flux today is $t_0$, the present age of the universe.

Neglecting cosmological effects, the expected UHE gamma ray flux, $j(E_0)$, at an observed energy $E_0$ is simply

$$j(E_0) = \frac{1}{4\pi} \lambda(E_0) \Phi_\gamma \,, \tag{26}$$

where $\lambda(E_0)$ is the absorption path length. For illustration, we shall use the QCD motivated gamma ray injection spectrum described by Eqs. (17), (18) and (19). At an energy $E_0$ for which $x = 2E_0/m_X \ll 1$, we get

$$\Phi_\gamma(E_0) \simeq \frac{dn_X}{dt_c} \frac{2}{m_X} (0.404) \left(\frac{2E_0}{m_X}\right)^{-1.5} \,. \tag{27}$$

Using Eqs. (15) and (16) for $dn_X/dt_c$, and taking the gamma ray path length from Ref. [43], we get, for $E_0 = 3 \times 10^{20}$ eV,

$$E_0^3 j(E_0) \simeq 7.3\times 10^{64}\left(\frac{dn^c_{M\bar{M}}}{dt_c}\cdot \mathrm{m}^3\,\mathrm{sec}\right) \times\left(\frac{\lambda}{10\,\mathrm{Mpc}}\right)\left(\frac{m_M}{m_X}\right)\left(\frac{m_X}{10^{15}\,\mathrm{GeV}}\right)^{1/2}\mathrm{eV}^2\,\mathrm{m}^{-2}\,\mathrm{sec}^{-1}\,\mathrm{sr}^{-1}\,. \quad (28)$$

Comparing this with the observed flux (corresponding to the Fly's Eye's observed highest energy event), which is $E_0^3 j(E_0 = 3\times 10^{20}\,\mathrm{eV}) \simeq 2\times 10^{25}\,\mathrm{eV}^2\,\mathrm{m}^{-2}\,\mathrm{sec}^{-1}\,\mathrm{sr}^{-1}$, we get

$$\frac{dn^c_{M\bar{M}}}{dt_c} \simeq 2.74\times 10^{-40}\left(\frac{10\,\mathrm{Mpc}}{\lambda}\right)\left(\frac{m_X}{m_M}\right)\left(\frac{10^{15}\,\mathrm{GeV}}{m_X}\right)^{1/2}\mathrm{m}^{-3}\,\mathrm{sec}^{-1}\,. \quad (29)$$

Now, using Eqs. (16), (29) and (25), and noting that

$$\left(\frac{1+z_c}{1+z_f}\right)^3 = \frac{n^c_M}{n^f_M}\,, \quad (30)$$

where the superscripts $c$ and $f$ refer to the quantities at the time of collapse and formation, respectively, we get, taking $t_i = t_c = t_0 = 2.057\times 10^{17} h^{-1}\,\mathrm{sec}$, the following expression for the required relative monopolonium abundance:

$$\xi^f \simeq 5.4\times 10^{-9}\left(\Omega_M h^2\right)^{-1} h^{-1}\left(\frac{m_X}{10^{15}\,\mathrm{GeV}}\right)^{1/2}\left(\frac{10\,\mathrm{Mpc}}{\lambda}\right)\left(\frac{422\,\mathrm{cm}^{-3}}{n_\gamma(t_0)}\right)\,. \quad (31)$$

In Eq. (31) $\Omega_M$ is the mass density contributed by monopoles (in units of the critical density of the universe), which is related to the ratio of monopole-to-photon number densities, $n_M/n_\gamma$, through the relation

$$\frac{n_M}{n_\gamma} = 2.49\times 10^{-24}\,\Omega_M h^2\left(\frac{m_M}{10^{16}\,\mathrm{GeV}}\right)^{-1}\,, \quad (32)$$

and $n_\gamma(t_0)$ is the photon number density in the universe today. We should perhaps mention that throughout this paper we assume the monopolonium formation rate to be low enough (which is justified a posteriori) that the ratio $n_M/n_\gamma$ is approximately constant.

Thus, Eq. (31) gives us a rough estimate of the required monopolonium abundance relative to any given monopole abundance. The results of our numerical calculations described in the next subsection confirm this expectation. As expected we need larger relative monopolonium abundance for smaller monopole abundance. On the other hand, for a given monopole number density, the larger the value of $m_X$ (or equivalently $m_M$), the larger is the flux produced (see, e.g., Eqs. (27) and (28)), and hence, smaller is the required relative monopolonium abundance (note that $\Omega_M$ in Eq. (31) is proportional to $m_X$).

### 3.5 Results

We now describe the results of our full numerical calculations of the flux and the required relative monopolonium abundance.

The recent results of the Fly's Eye experiment [7] indicate a change of composition of the UHE cosmic rays from a predominantly heavy component at energies below about $2\times10^{18}$ eV to a predominantly light component above that energy. Further, this composition change is correlated with a dip in the overall spectrum [7]. The Fly's Eye group obtained a good fit to the observed spectrum by superposing a steeper galactic component of iron nuclei and a flatter extragalactic component of protons. If we assume that the extragalactic component is dominated by monopole annihilation at the high energy end we can determine the necessary relative monopolonium abundance mentioned above by normalizing the calculated flux to this light component. In the Fly's Eye as well as in the AGASA data this high energy end which we shall use as our normalization point is located at about $E_0 = 10^{19.7}$ eV. Between this energy and the respective highest energies measured there is a gap in both data sets which could indicate a break in the spectrum. It will be clear from Figs. 1 and 2 presented below that, because of the specific shapes of the spectra, normalization at a lower energy would be inconsistent because then the calculated flux would exceed the observed flux at a higher energy. In this sense, the above normalization procedure is the only consistent normalization in our case. Note also that since gamma rays and nucleons can not be distinguished by the Fly's Eye at these energies we have to use for normalization the combined nucleon and gamma ray flux given by our model.

In our numerical calculations we use $\alpha_m = 34.25$ from Eq. (5) and $\alpha_X = 1/30$. The monopole abundance $\Omega_M$ is constrained by various bounds [33]. Currently, the most stringent (and theoretically well-motivated) constraint on $\Omega_M$ is that given by a recent modification [37] of the original Parker bound [36], giving $\Omega_M h^2 \leq 3.99 \times 10^{-3} \left(m_M/10^{16}\,\mathrm{GeV}\right)^2$ for monopole velocities $\sim 10^{-3}c$. (Note that current *experimental* upper limits [35] on $\Omega_M$ are still about a factor of 5 weaker than the recent Parker bound). For reference we will therefore use the improved Parker bound [37] for $\Omega_M$. The monopole mass is fixed by $m_X$ and $\alpha_X$. We will show the results for two values of $m_X$, namely, $m_X = 10^{15}$ GeV and $10^{16}$ GeV.

Figs. 1 and 2 show the nucleon, gamma ray and neutrino spectra obtained by the normalization procedure mentioned above. Figs. 1 and 2 correspond to the QCD motivated and the phenomenological injection spectra described in section 3.2, respectively. As expected, the gamma rays and neutrinos far outnumber the protons at energies above $\sim 10^{20}$ eV.

We can compare the predicted and the "measured" fluxes at the highest energies in the following somewhat more quantitative way: The integral CR flux above $1.7\times10^{20}$ eV estimated from the highest energy Fly's Eye and AGASA events is $\simeq 0.7 \times 10^{-16}\,\mathrm{m}^{-2}\,\mathrm{sec}^{-1}\,\mathrm{sr}^{-1}$ [51]. The corresponding gamma ray flux expected within our model is about a factor 10 (for QCD motivated injection, see Figs. 1) and 3 (for phenomenological injection, see Figs. 2) larger. In contrast, the expected nucleon fluxes are smaller by about a factor 3 and 5, respectively. Using the additional information about the event energies and comparing with the predicted differential flux (see error bars in Figs. 1 and 2) our model is at least for the phenomenological injection spectrum in good agreement with gamma rays as primaries for the two highest energy events whereas nucleons seem to be disfavored.

Taking the enhancement factor due to neutrino cascading mentioned in section 3.3 into account the predicted differential neutrino flux is between one and two orders of magnitude smaller than current limits on the flux of deeply penetrating particles from the Fly's Eye experiment [52, 53]. Furthermore, the predicted integral neutrino flux above $\simeq 10^{20}\,$eV is smaller than current limits from the Fréjus detector [54] by factors of $\simeq 10^{-2}$ (QCD motivated injection) and $\simeq 10^{-4}$ (phenomenological injection).

Finally, Figs. 3 and 4 show the contours of the required values of $\xi$, the relative monopolonium abundance (or equivalently, the value of the parameter $\eta$, see section 4 below), obtained by the normalization procedure described above, in the $\Omega_M - m_X$ plane. The shaded regions are excluded by the Parker bound on $\Omega_M$. The sharp upturn of the curves for very low values of $m_X$ in Figs. 3 and 4 is simply a reflection of the fact that for too low values of $m_X$ the maximum available energy $\sim m_X/2$ becomes lower than the energy of the cosmic rays under consideration.

# 4 Equilibrium Estimate of Monopolonium Abundance and the Resulting Cosmic Ray Flux

In this section, we estimate the relative monopolonium abundance in the universe by using the classical differential version of the Saha equation applied to a system of $M$s, $\bar{M}$s and $M-\bar{M}$ bound states in thermal equilibrium with the background relativistic plasma (see the discussions at the end of section 2). Following Hill [32] we shall use the Maxwell-Boltzmann distribution for the number density of monopoles to estimate the number density of monopolonia. As Hill pointed out, the equilibrium analysis is expected to yield a *lower limit* to the true monopolonium abundance, i.e., consideration of any specific mechanism of monopolonium formation is likely to *increase* the monopolonium abundance, which would in the end increase the contribution to the total cosmic ray flux. In other words, the equilibrium monopolonium abundance estimated below probably gives a conservative estimate of the possible contribution of monopolonia to the cosmic ray flux.

Within the framework of an equilibrium analysis, the operational definition of monopolonium "formation" adopted in section 2 implies that at a time $t$ when the temperature of the universe is $T$, monopolonia with binding energy $E_b = \eta T$ are "formed" with an abundance equal to the equilibrium abundance of $M - \bar{M}$ bound states of binding energy $\eta T$. The equilibrium abundance can be calculated by using the Saha formalism. Thus monopolonia formed between times $t_f$ and $t_f + dt_f$ have binding energies between $E_{bf}$ and $E_{bf} + dE_{bf}$. Their number density can be written as [32]

$$dn^f_{M\bar{M}} = \frac{dE_{bf}}{E_{bf}}\left(\frac{\pi^3}{2}\right)\alpha_m^3\left(\frac{1}{2\pi T_f E_{bf}}\right)^{3/2} n_M n_{\bar{M}} e^{|E_{bf}|/kT_f}\,, \tag{33}$$

where $E_{bf} = \eta T_f$, and $n_M$ ($n_{\bar{M}}$) denotes the number density of monopoles (antimonopoles) at the time $t_f$. Using $n_\gamma = 2\zeta(3)T_f^3/\pi^2$ ($\zeta(3) = 1.202$) for the photon density at temperature

$T_f$ and Eq. (32) we can write Eq. (33) in the form

$$\frac{dn^f_{M\bar{M}}}{n_M} = 5.97 \times 10^{-25} \frac{dE_{bf}}{E_{bf}} \alpha_m^3 \eta^{-3/2} e^{\eta} \Omega_M h^2 \left(\frac{m_M}{10^{16}\,\mathrm{GeV}}\right)^{-1} . \tag{34}$$

We thus see that the quantity $\eta$ parametrizes the bound state abundance for a given monopole abundance $\Omega_M$. The relative monopolonium abundance at formation, $\xi_f$, is then given by Eq. (25).

Using Eqs. (15), (16) and (34), and converting redshifts to time by using standard cosmological relations [33], we can write

$$\begin{aligned} \frac{dn_X}{dt_i} &= \frac{4\zeta(3)^2}{3\pi} T_0^3 \alpha_m^3 \left(\frac{m_M}{m_X}\right) \frac{e^{\eta}}{(2\pi\eta)^{3/2}} \left(\frac{n_M}{n_\gamma}\right)^2 t_0^2 t_i^{-3} \\ &= 1.32 \times 10^{-57}\, \alpha_m^3 \left(\frac{m_M}{m_X}\right) \left(\Omega_M h^2\right)^2 \left(\frac{m_M}{10^{16}\,\mathrm{GeV}}\right)^{-2} \eta^{-3/2} e^{\eta} \\ &\quad \times \left(\frac{10^{10}\mathrm{yr}}{t_0}\right) \left(\frac{T_0}{2.75\mathrm{K}}\right)^3 \left(\frac{t_0}{t_i}\right)^3 \mathrm{m}^{-3}\,\mathrm{sec}^{-1} , \end{aligned} \tag{35}$$

where $T_0$ is the photon temperature today and $t_0$ is the present age of the universe.

Introducing the generic form [19]

$$\frac{dn_X(t_i)}{dt_i} = \kappa m_X^p t_i^{-4+p} , \tag{36}$$

where $\kappa$ and $p$ are dimensionless constants whose values depend on the specific process involving the specific kind of TD under consideration we see that Eq. (35) corresponds to $p = 1$ and

$$\kappa = 7.38 \times 10^{-20} \alpha_m^3 \left(\frac{m_M}{10^{16}\,\mathrm{GeV}}\right)^{-3} \left(\frac{m_M}{m_X}\right)^2 \left(\Omega_M h^2\right)^2 \eta^{-3/2} e^{\eta} \left(\frac{T_0}{2.75\mathrm{K}}\right)^3 \left(\frac{t_0}{10^{10}\mathrm{yr}}\right)^2 . \tag{37}$$

Using Eqs. (17), (35) and (23), we finally get the following expression for the expected flux in terms of the parameter $\eta$:

$$\begin{aligned} E_0^3 j_a(E_0) &= 3 \times 10^3\, \alpha_m^3 \left(\frac{10^{16}\,\mathrm{GeV}}{m_M}\right)^3 \left(\frac{m_M}{m_X}\right)^2 (\Omega_M h^2)^2 \eta^{-3/2} e^{\eta} \left(\frac{T_0}{2.75\,\mathrm{K}}\right)^3 \\ &\quad \times \left(\frac{E_0}{10^{20}\,\mathrm{eV}}\right)^3 \int_0^\infty \frac{1}{1+z_i} \frac{dE_i(E_0, z_i)}{dE_0} \frac{dN_a}{dx}\, \mathrm{eV}^2\,\mathrm{m}^{-2}\,\mathrm{sec}^{-1}\,\mathrm{sr}^{-1} , \end{aligned} \tag{38}$$

where $x \equiv 2E_i/m_X$.

Figs. 1 and 2 show the expected flux of various particles for various different injection spectra and for different values of $m_X$. The required value of the parameter $\eta$ is obtained by the normalization procedure described in section 3.5. Figs. 3 and 4 show the contour plots for the normalized value of $\eta$ in the $\Omega_M - m_X$ parameter space. From these plots we see that even if the actual value of $\Omega_M$ is significantly less than the Parker bound value, collapse of monopolonia can give significant contribution to UHE CR provided $\eta$ is sufficiently large (say, between 30 and 70). Indeed, as is clear from Eq. (38), the contribution of monopolonia to the flux of cosmic rays increases exponentially with the parameter $\eta$. Of course, for too large values of $\eta$, the monopolonium is so tightly bound that the monopole and the antimonopole constituting it do not interact with other monopoles/antimonopoles but are dominated by the local mutual coulomb interaction. As a consequence, the equilibrium analysis is not expected to be valid for large $\eta$.

# 5 Conclusions

In summary, we have studied the process of monopolonium collapse as a possible source of production of the UHE cosmic rays. The properties of the spectra of the UHE nucleons, photons and neutrinos produced by this process are essentially same as those already described in earlier works [19, 43] that dealt with UHE cosmic ray production from topological defects in general; the reader is advised to consult these references for details of the spectra. In the present paper, our main aim has been to estimate the overall magnitude of the flux of UHE cosmic rays produced by monopole annihilations. This is determined not only by the actual abundance of monopoles in the universe (which, unfortunately, is unknown), but also by the monopolonium abundance relative to monopoles. We have estimated the relative monopolonium abundance that would be required in order to produce sufficient flux of the UHE cosmic rays as observed. For a monopole abundance saturating the Parker bound, say, a fractional monopolonium abundance of $\sim 10^{-8}$ (with $m_X \sim 10^{16}$ GeV) can give rise to measurable UHE CR flux and can, in principle, explain the recent HECR events [8, 10]. The required fractional monopolonium abundance, however, increases with decreasing value of the monopole abundance. Within the context of an analysis that assumes monopoles to be in thermal equilibrium at the relevant time of monopolonium formation, we find that the required relative monopolonium abundances may well be realizable. Out of equilibrium processes like transition to a transient superconducting phase [55, 56] giving rise to transient magnetic flux tubes connecting monopole-antimonopole pairs could well enhance monopolonium abundances beyond the equilibrium values and compatible with the required numbers derived above. A full microscopic treatment is, however, required in order to calculate these abundances from first principles. Work along this direction is in progress.

# Acknowledgments

We thank Chris Hill and David Schramm for discussions, constant encouragement and support. We would also like to thank Tom Kibble, Albert Stebbins, Neil Turok, Alex Vilenkin and Alan Watson for useful discussions. Furthermore, we are grateful to the Fly's Eye group, especially Hongyue Dai, Paul Sommers and Shigeru Yoshida, for discussions of various aspects of the current experimental status of UHE CR observations. This work was supported by the DoE, NSF and NASA at the University of Chicago, by the DoE and by NASA through grant NAGW-2381 at Fermilab, and by the Alexander-von-Humboldt Foundation. One of us (PB) would like to thank the Isaac Newton Institute, Cambridge, UK., where a major part of this work was done, for hospitality and financial support.

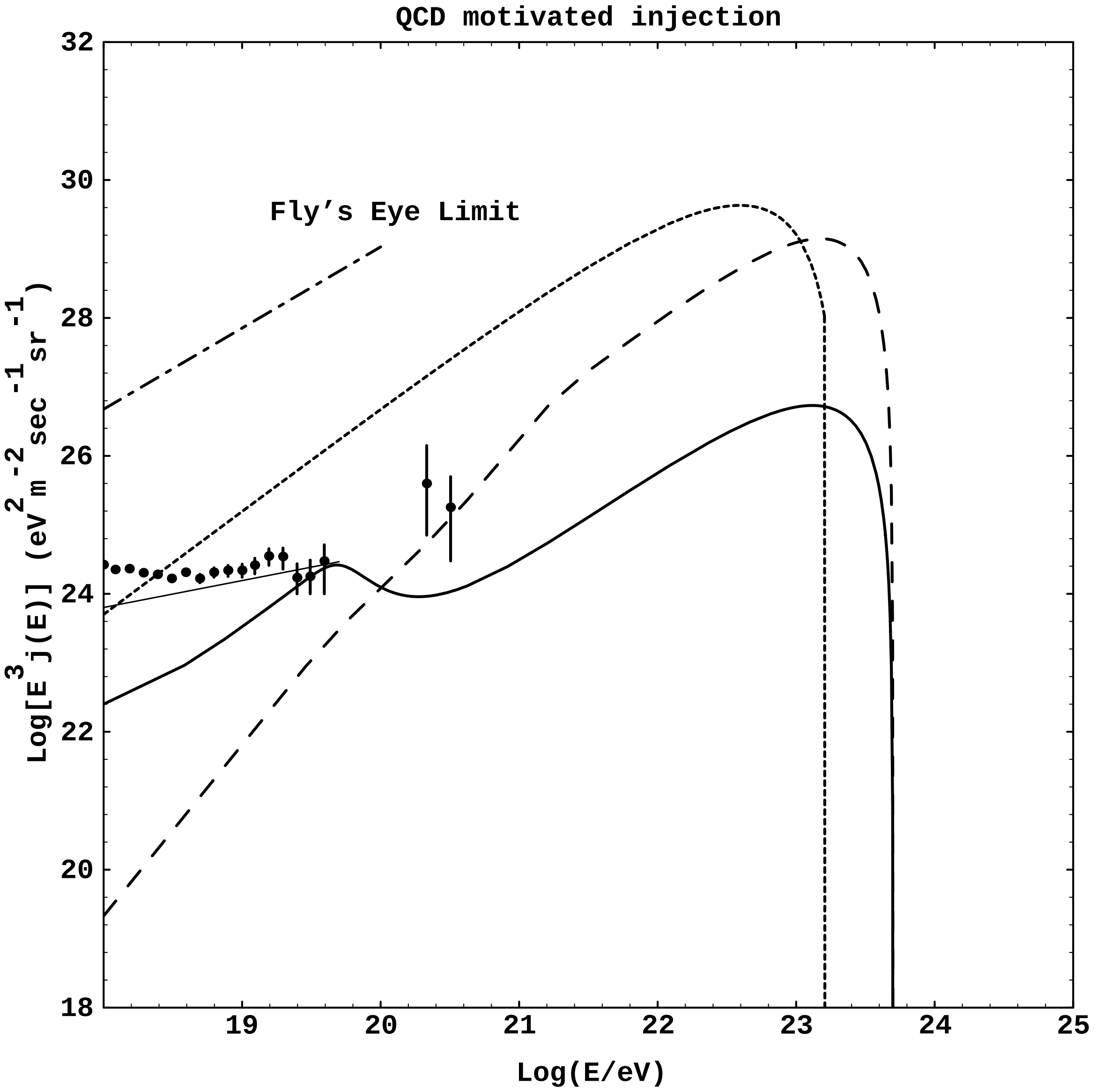

**Figure 1A:** Observable neutrino (short-dashed line), gamma ray (long-dashed line) and proton (solid line) spectrum produced by monopolonium collapse. An X particle mass of $m_X = 10^{15}$ GeV and the QCD-motivated injection spectra discussed in section 3.2 were used. The combined proton and gamma ray flux was normalized at $10^{19.7}$ eV to the "extragalactic flux component" (thin solid line) [7] fitted to the high energy Fly's Eye data (dots with error bars) as described in section 3.5. Also shown (dash-dotted line) is an approximate limit on the neutrino flux determined from the non-detection of deeply penetrating particles by the Fly's Eye detector [52, 53].

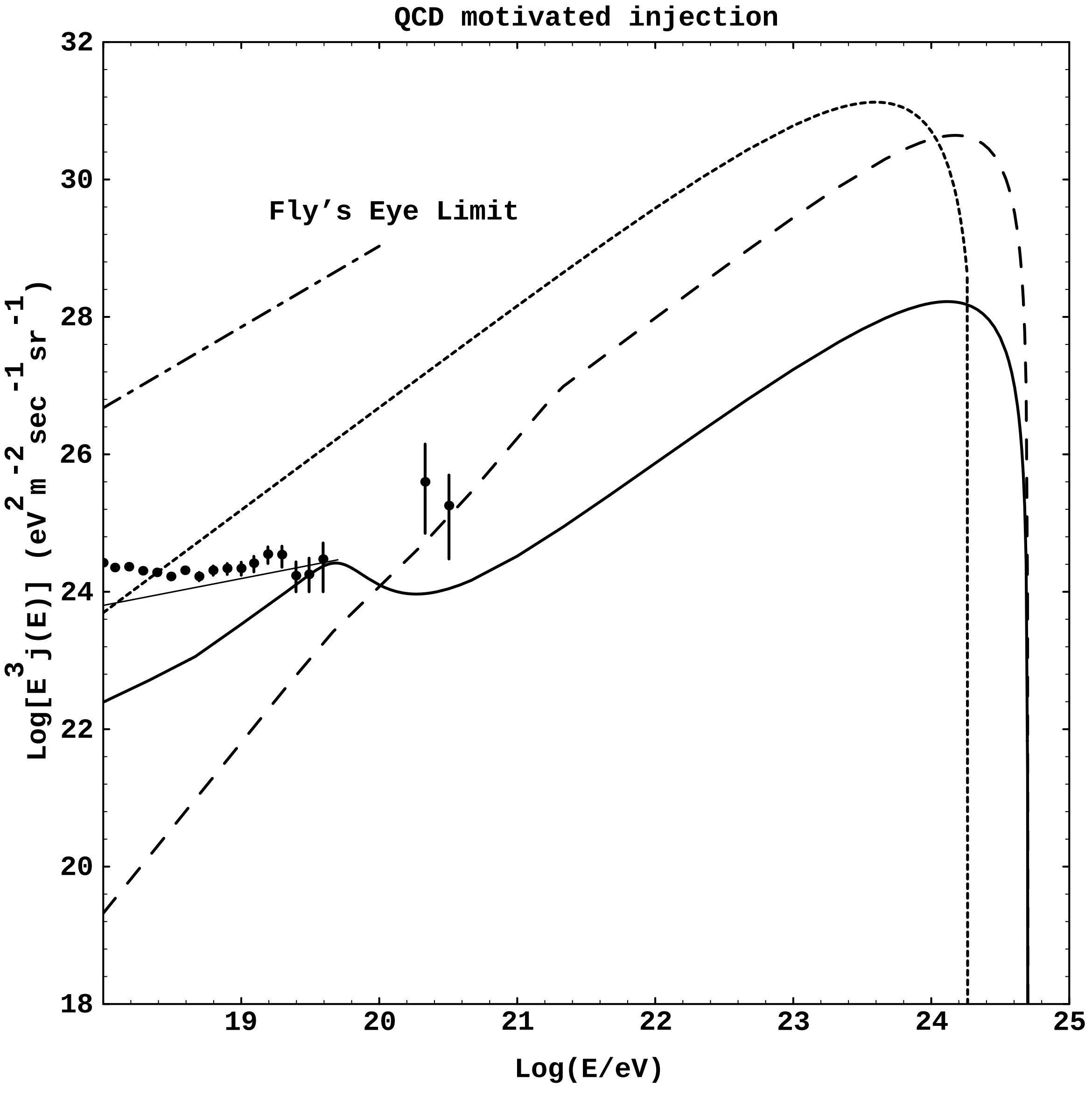

**Figure 1B:** Same as Fig. 1A but for an X particle mass $m_X = 10^{16}$ GeV.

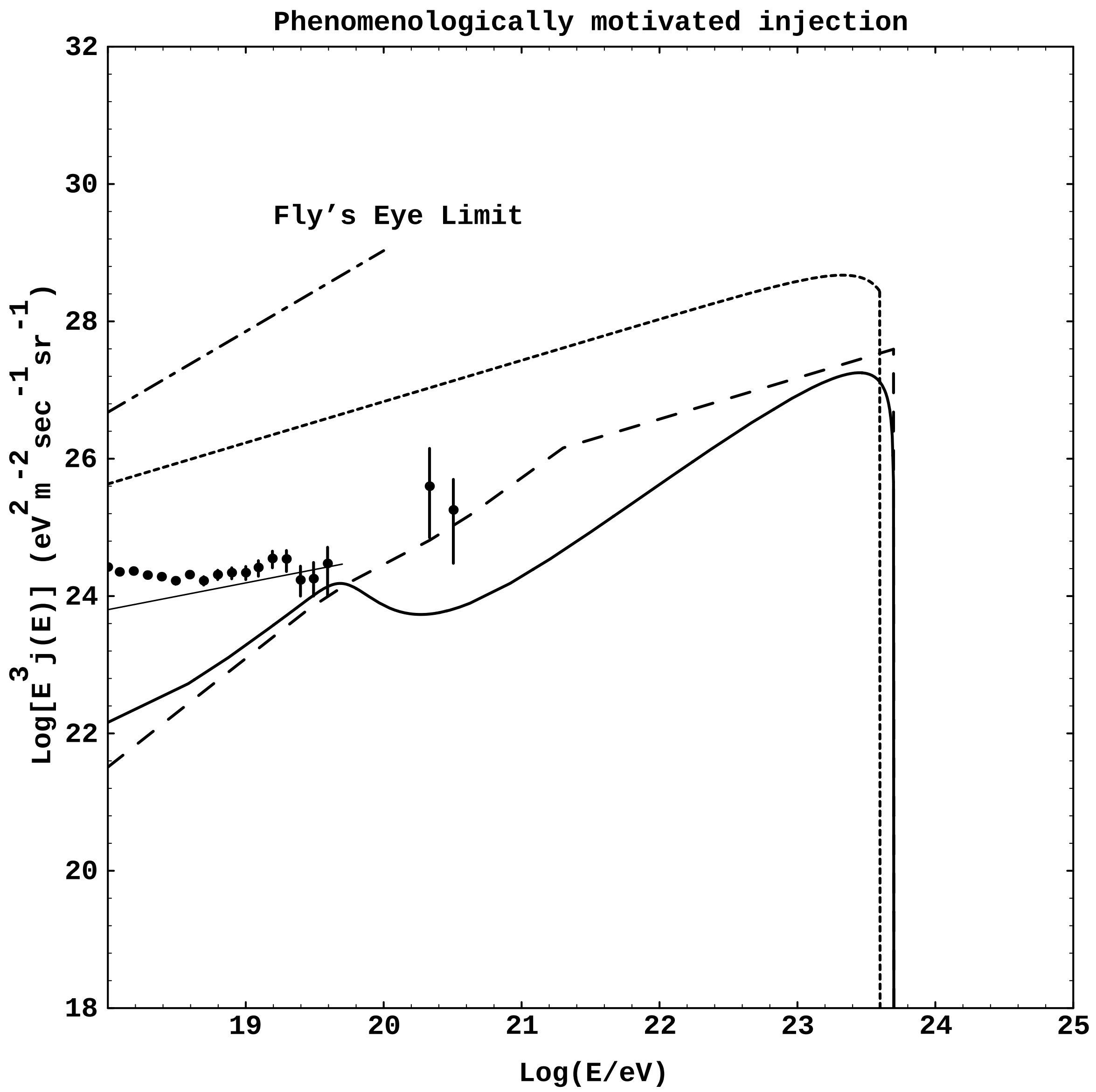

**Figure 2A:** Same as Fig. 1A, except that the phenomenologically motivated injection spectra discussed in section 3.2 were used. Edges in the gamma ray and neutrino spectrum are due to simplifying numerical assumptions.

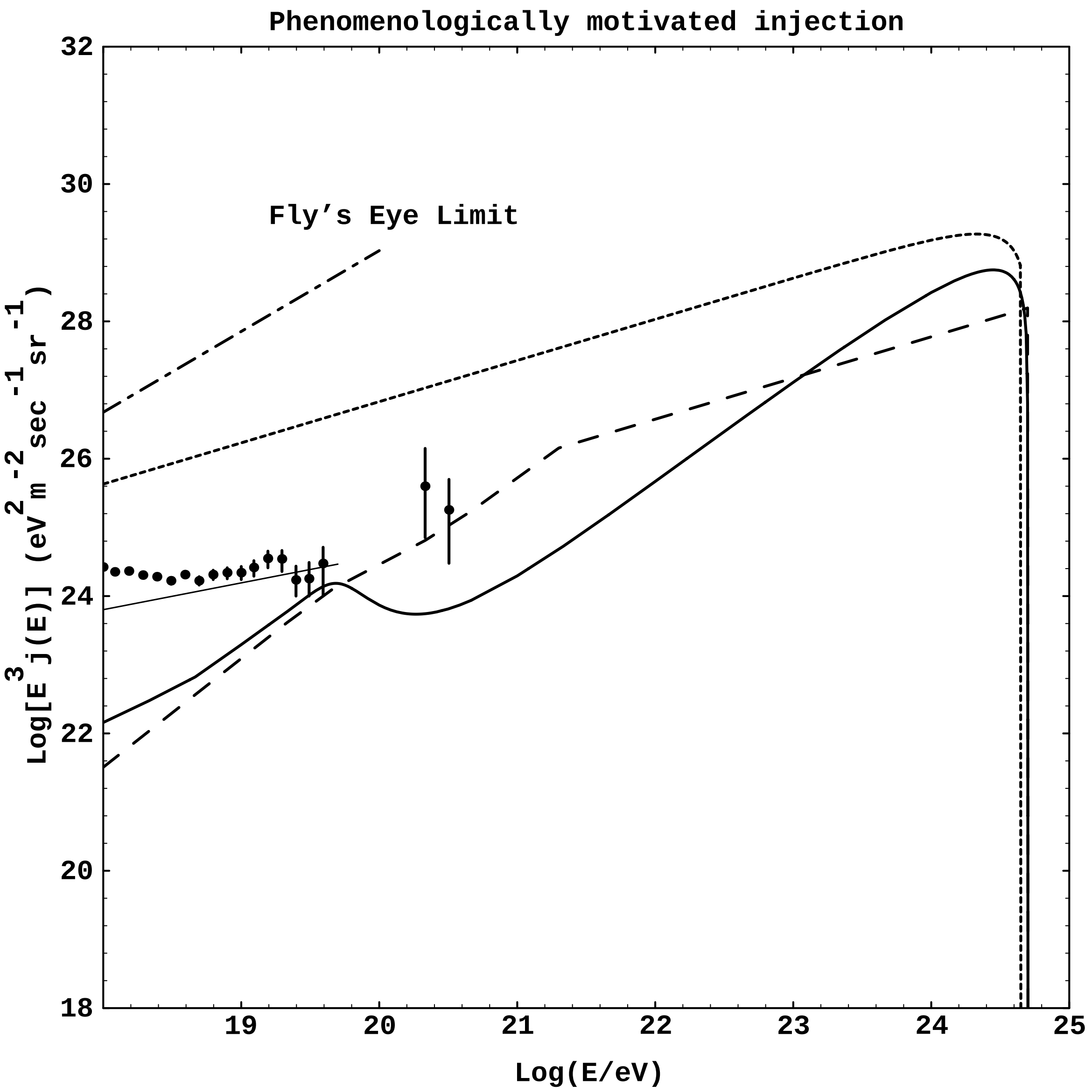

**Figure 2B:** Same as Fig. 1B, except that the phenomenologically motivated injection spectra discussed in section 3.2 were used. Edges in the gamma ray and neutrino spectrum are due to simplifying numerical assumptions.

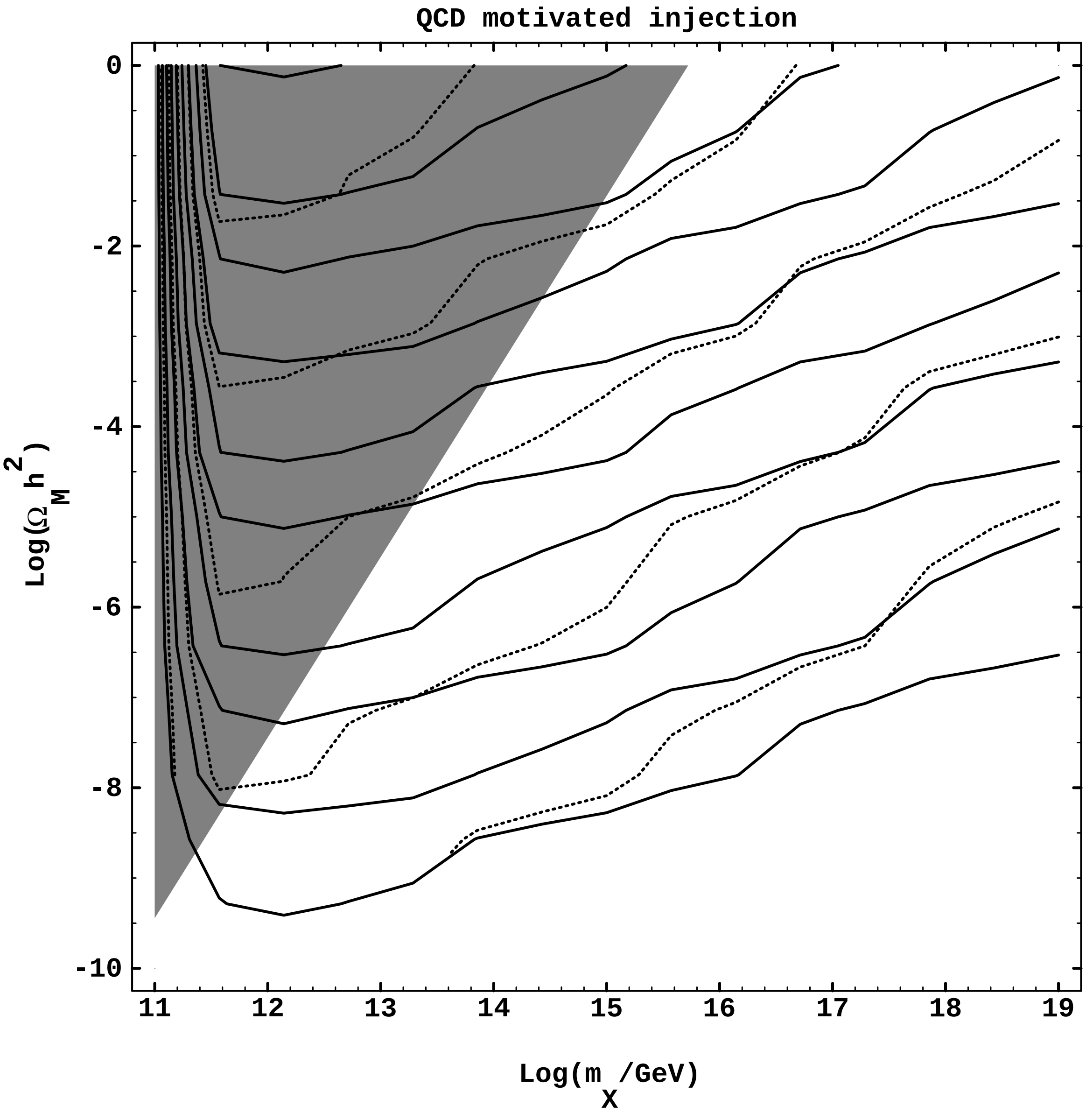

**Figure 3:** Contours of the relative monopolonium abundance $\xi_f$ as defined in Eq. (25) (solid lines) and the corresponding $\eta$-parameter within the equilibrium analysis (dashed lines) following from the normalization explained in section 3.5. QCD-motivated injection as discussed in section 3.2. was assumed. In the $m_X - \Omega_M h^2$ plane the $\xi_f$- and $\eta$-contour lines correspond to $1, 0.1, \cdots, 10^{-9}$ and $70, 60, \cdots, 30$, respectively, decreasing from the lower right to the upper left. The shaded parameter region is excluded within the modified Parker bound [37].

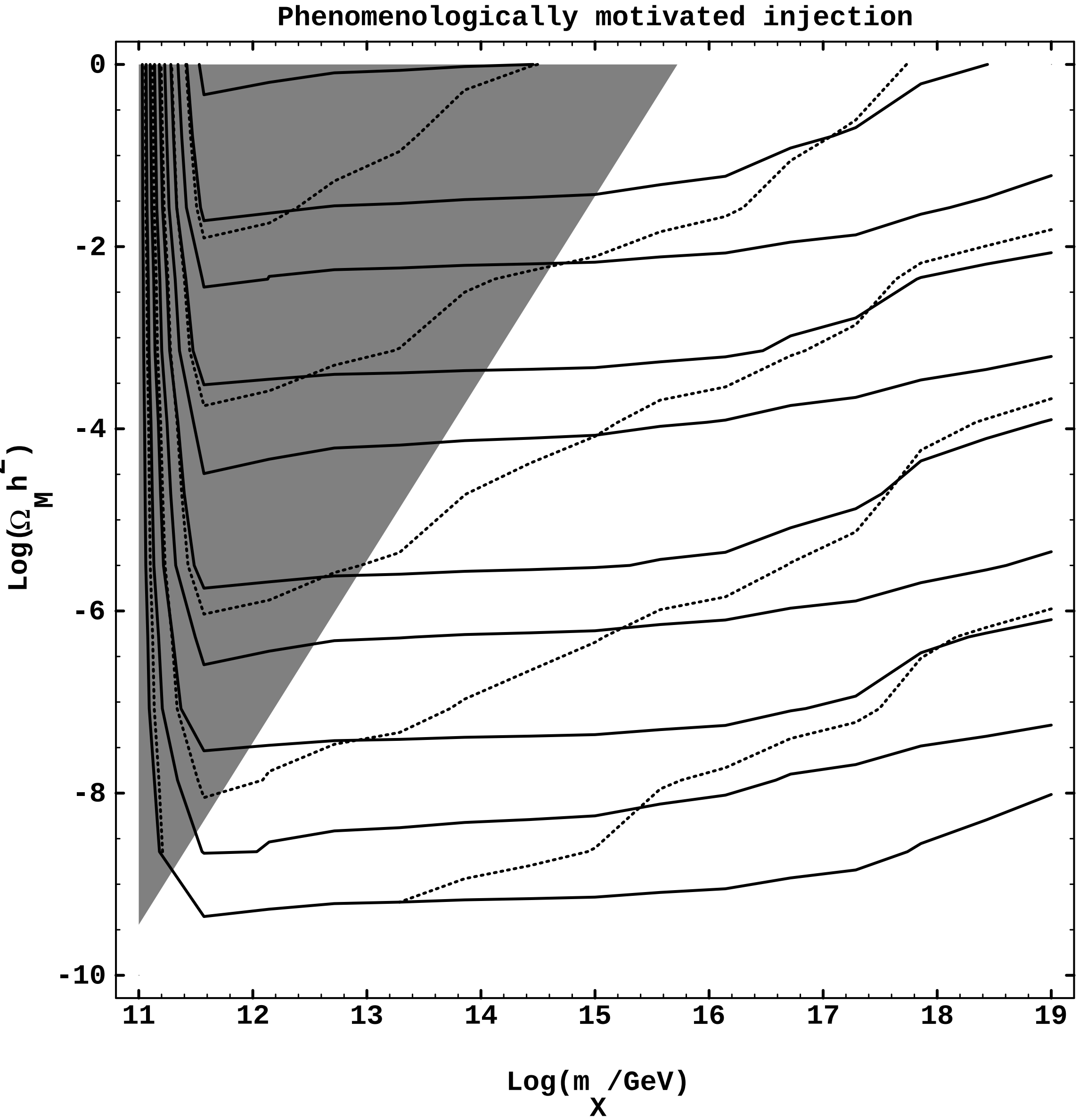

**Figure 4:** Same as Fig. 3 but assuming the phenomenological injection spectra discussed in section 3.2.